\documentclass[twocolumn, superscriptaddress]{revtex4}
\pdfoutput=1
\usepackage{epsfig,graphicx}
\usepackage{amsmath,amssymb,amsfonts,amsthm}
\usepackage{bm} 
\usepackage{flafter}
\usepackage{color}
\usepackage{epstopdf}
\usepackage{latexsym}
\usepackage[colorlinks=true, citecolor=blue, urlcolor=blue ]{hyperref}
\usepackage{pxfonts,txfonts}
\usepackage{tgpagella}
\usepackage[T1]{fontenc}
\usepackage{ulem} 
\usepackage[utf8]{inputenc}
\newcommand{\bra}[1]{\langle #1 |}
\newcommand{\ket}[1]{| #1 \rangle}

\DeclareMathOperator{\tr}{Tr}

\newcommand{\la}[1]{\lambda}
\graphicspath{{figures/}}
\usepackage{braket}
\graphicspath {{figures/}}

\usepackage{float}

\begin{document}

	\title{Frustrated quantum spin systems in small triangular\\ lattices studied with a numerical method}
	
\author{D. Castells-Graells}
\affiliation{ F\'{\i}sica Te\`{o}rica: Informaci\'{o} i Fen\`{o}mens Qu\`{a}ntics, Departament de F\'{\i}sica, Universitat Aut\`{o}noma de Barcelona, 08193 Bellaterra, Spain}
\affiliation{ Physics Department, ETH Z\"urich, CH-8093 Z\"urich, Switzerland}
\author{A. Yuste}
\affiliation{ F\'{\i}sica Te\`{o}rica: Informaci\'{o} i Fen\`{o}mens Qu\`{a}ntics, Departament de F\'{\i}sica, Universitat Aut\`{o}noma de Barcelona, 08193 Bellaterra, Spain}
\affiliation{ IBM, Avinguda Diagonal, 571, 08029 Barcelona, Spain}
\author{A. Sanpera}
\email{Anna.Sanpera@uab.cat} 
\affiliation{ F\'{\i}sica Te\`{o}rica: Informaci\'{o} i Fen\`{o}mens Qu\`{a}ntics, Departament de F\'{\i}sica, Universitat Aut\`{o}noma de Barcelona, 08193 Bellaterra, Spain}
\affiliation{ICREA, Psg. Llu\'is Companys 23, 08010, Barcelona, Spain.}

\begin{abstract}
The study of quantum frustrated systems remains one of the most challenging subjects of quantum magnetism, as they can hold quantum spin liquids, whose characterization is quite elusive.  The presence of gapped quantum spin liquids possessing long range entanglement while being locally indistinguishable often demand highly sophisticated numerical approaches for their description. Here we propose an easy computational method based on exact diagonalization with \textit{engineered} boundary conditions in very small plaquettes. We apply the method to study the quantum phase diagram of diverse antiferromagnetic frustrated Heisenberg models in the triangular lattice.  Our results are in qualitative agreement with previous results obtained by means of sophisticated methods like 2D-DMRG or variational quantum Monte Carlo.	
\end{abstract}

\maketitle
\section{Introduction}

Some entangled ground states of spin systems do not order even at zero temperature. The lack of order, which is originated by strong quantum fluctuations on the spin orientations, prevents their characterization by means of local order parameters. Such quantum disordered states, termed generically quantum spin liquids (QSL), are linked to topological degenerated ground states and contain long range entanglement. Moreover, they are locally indistinguishable~\cite{Wen2002, balents2010, Savary2017}, meaning that they cannot be detected or distinguished using local measurements.

QSL are often caricatured as a liquid of singlets, where the singlets formed between nearby spins strongly fluctuate from one configuration to another. Due to such fluctuations, the ground state of the system is far from a product state, implying that entanglement in QSL plays a crucial role. Ground states of local spin Hamiltonians are normally short range entangled, as evidenced by the fact that the entanglement entropy, $\mathcal{S}$, of any bipartite cut of the system follows an area law: $\mathcal{S}(L)\sim L^{D-1}$, where $D$ is the dimension of the system and $L$ the linear size of the boundary separating both regions. Corrections to this law appear, for instance, in critical gapless quantum phases or in topologically ordered states. In 2D, the latter fulfill $\mathcal{S}(L) \sim L+ b_0\gamma$, where $\gamma$ is a universal correction called topological entanglement entropy, which is independent of the lattice size an signals topological order~\cite{Kitaev2006,Levin2006,Isakov2011}.

The combination of the above features makes unfeasible to use effective mean field approaches with fluctuation corrections over the mean field ansatz for the description of QSL. Hence, finding for such cases the eigenstates of the Hamiltonians of interest mostly relies, for the time being, in numerical approaches and/or complex variational ansatzs. The numerical methods are, of course, severely hindered by the requirement of large lattices.

Although exact diagonalization (ED) methods suffer from strict size constrains, which can be slightly leveraged when symmetries are cleverly implemented, here we approach the study of quantum frustrated systems by using Lanczos-based ED in very small system sizes at the expense of  properly engineering the boundary conditions.  With this method, we search for signatures of putative gapped QSL models in frustrated systems. On the one hand, we are able to reproduce in a good qualitative agreement the quantum phase diagram of some paradigmatic frustrated models that have been previously reported in the literature. On the other hand, we exploit our method to investigate unexplored frustrated models. Interestingly enough, some signatures of gapped QSL as, for instance, the lack of an order parameter, the topological degeneracy, the increase of entanglement or the blurring of defined peaks in the spin structure factor can be observed using properly engineered boundary conditions.

 We focus our analysis to spin-1/2 antiferromagnetic (AF) Heisenberg models in the triangular lattice, a paradigmatic geometry where quantum fluctuations and frustration compete. The effect of frustration, i.e. the impossibility to simultaneously minimize the Hamiltonian locally, can be further tuned if the couplings along different lattice directions are anisotropic. In this context, the paradigmatic model is the Heisenberg model with spatial anisotropy between horizontal and diagonal bonds, the so-called SATL model. Such model has been extensively addressed in the literature using different methods such as tensor networks, quantum Monte Carlo, 2D DMRG, exact diagonalization (ED) or modified spin wave theory (MSWT), see e.g. \cite{Leung93,Bernu94,Yunoki2006, Weng2006, Schmied2008, Tocchio2009, Heidarian2009, Hauke2010, Weichselbaum2011,PhysRevB.83.100405, Tocchio2013,doi:10.7566/JPSJ.86.114705, Thesberg2014,Celi2016,Yuste2017,PhysRevB.95.165110}, all of them reporting the existence of gapped QSL in some regions of the phase-space diagram.  Here, we consider the straight generalization of the SATL model; the spatially {\textit{completely}} anisotropic triangular lattice (SCATL) with anisotropic couplings along {\textit{all}} lattice directions. It is important to remark that the SCATL model has been scarcely addressed in the literature. We also investigate here the  $J_1$-$J_2$ model in the triangular lattice, where the anisotropy is now introduced between the nearest-neighbor ($J_1$) and next-to-nearest neighbor ($J_2$) couplings. The presence of gapped QSL in this model has also been addressed recently in the literature \cite{Zhu2015,Hu2015,Iqbal2016,Hu2016,Wietek2017,Gong2017}. All the above models give room to both gapless and highly nontrivial gapped QSL. Finally, we propose and study a hybrid model between the anisotropic SATL and the $J_1$-$J_2$ model, which we denote as  the "anisotropic $J_1$-$J_2$ model". Such a hybrid model reduces to the SATL in the limit $J_2\rightarrow 0$, and to the standard  $J_1$-$J_2$ model in the limit where the anisotropy between horizontal and diagonal bonds disappears.  Our aim, asides of gaining further insight in frustrated models, is to investigate if the predicted QSL present in the SATL and $J_1$-$J_2$ Heisenberg models are connected and have, therefore, the same nature.

Before proceeding further, we summarize our main results. We derive a quantum phase diagram for the above models using ED with engineered boundary conditions in lattices of $N=9$, $12$ or $16$ spins. Our results reproduce quite closely both the ordered and disordered quantum phases previously reported. Our method relies on the fact that in the small lattice limit ordered phases correspond to precisely fixed boundary conditions, while there exist regions on the phase-space diagram where a massive number of different boundary conditions provide ground states whose energy is approximately equal. These regions match qualitatively the parameters for which gapped QSL have been previously predicted using MSWT, 2D-DMRG or PEPS. Our lattice sizes are definitively too small to show non-trivial topological invariants or the presence of topological entanglement entropy, but the calculation of the geometric entanglement  --quantifying how far an entangled state is from its closest separable one-- shows that these presumed gapped QSL phases have a large entanglement as compared to their surrounding ordered phases. Moreover, even with such small lattice sizes, it is possible to see in these regions of the phase space the presence of a topological degeneracy if the system is subjected to the effect of an external artificial magnetic flux.

The paper is organized as follows: in Sec.~\ref{Sec:RBC}, we explain the main features of our numerical method together with the relevant figures of merit used along. In Sec.~\ref{Sec:SCATL}, we derive the quantum phase diagram of the SCATL model with anisotropic couplings along all lattice directions. For this model, to the best of our knowledge, only a study based on a MSWT exists \cite{Hauke2010}. Therefore, alternative methods are clearly needed to settle the presence of conjectured QSL. In Sec.~\ref{Sec:J1J2}, we move onto another paradigmatic frustrated model, the so-called $J_1$-$J_2$. We analyze it also in the presence of chiral interactions, which helps to elucidate the nature of the predicted QSL. There, we compare our results with the quantum phase diagram obtained recently in \cite{Gong2017} using 2D DMRG. 
In Sec.~\ref{Sec:Hybrid}, we introduce the \textit{anisotropic $J_1$-$J_2$} Heisenberg model aiming at investigating the connection between the gapped QSL appearing in the SATL model with the ones appearing for the $J_1$-$J_2$ model. Finally, in Sec.~\ref{Sec:conclusions}, we conclude and present some open questions.

\section{Random Twisted Boundary Conditions}
\label{Sec:RBC}
\begin{figure*}[t]
	\includegraphics[width=1.4\columnwidth]{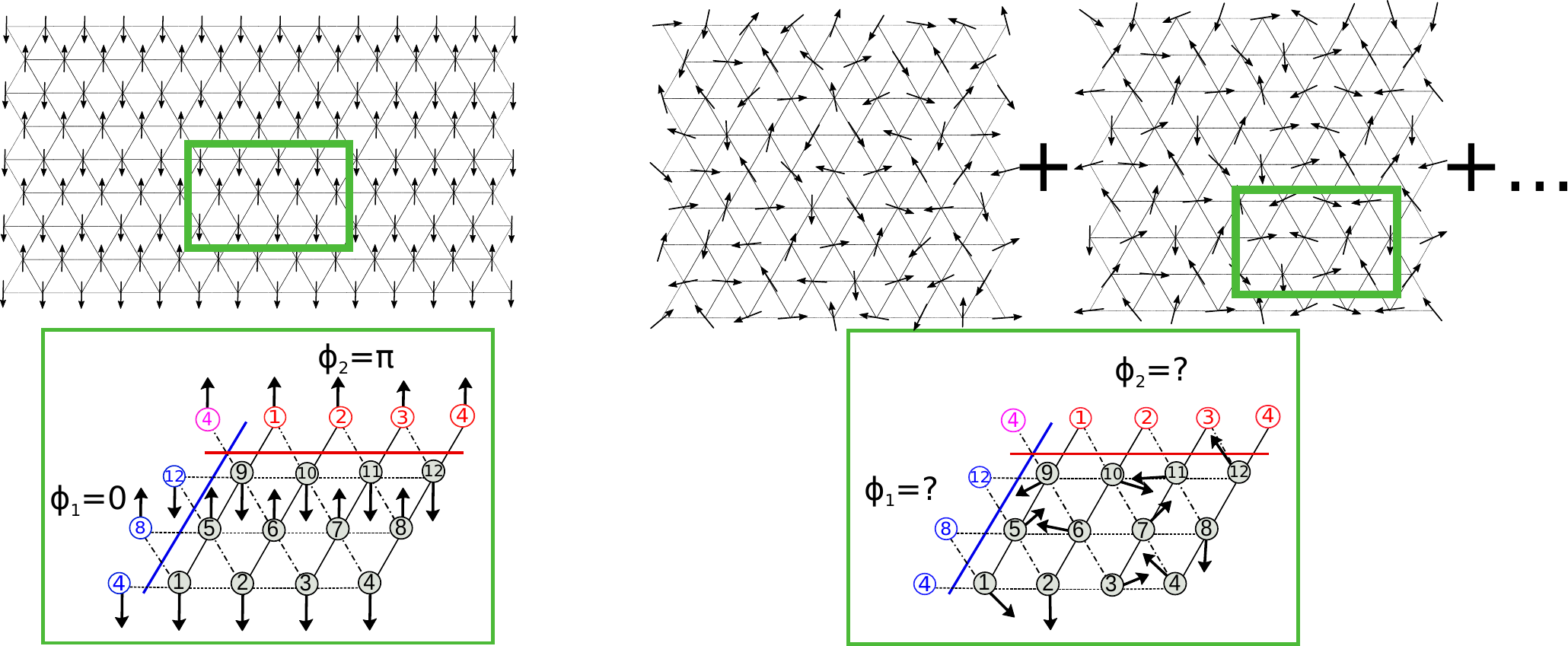}
	\caption{Upper panels: sketch representation of a quantum ordered N\'eel phase (left) and a QSL phase (right) in a large lattice. Lower panels: sketch of twisted boundary conditions in a $4\times3$ triangular lattice with anisotropic nearest-neighbor interactions. Boundary spins in blue are twisted in the XY plane by an angle $\phi_1$, while red colored boundary spins are twisted by a phase $\phi_2$.
		The pink colored boundary spin (top-left corner) is twisted by an angle $\phi=\phi_1+\phi_2$. Bottom left: for an ordered 2D-N\'eel phase along the diagonal directions, the boundary phases which reproduce the order are $\phi_1=0$ and $\phi_2=\pi$. Bottom right: For a quantum disordered phase, such a set of phases cannot be defined. The anisotropy of the SCATL model is depicted by the three different line styles in the bottom panels.}
	\label{fig:Lattice}
\end{figure*}
Twisted boundary conditions (TBC) were introduced in the seminal contributions of \cite{Poiblanc91,Gros92}, and can be thought of as periodic boundary conditions (PBC) under a twist. Since then, they have been often used to calculate properties of quantum magnets,
as they provide better access to momentum space and help to mitigate finite size effects, see e.g.~\cite{DeVries2009,Varney2011,Varney2012,Thesberg2014,Mendes-Santos2014}. More recently, periodic TBC in a 2D lattice have also been used to calculate the Chern numbers for many-body systems in a lattice \cite{Fukui2005} (playing the role of momenta $k_x$, $k_y$), and to investigate as well the topological degeneracy \cite{Thouless82,Nielsen2013} present in some chiral phases (see e.g. ~\cite{Haldane95, Neupert2011}).  However, here we use random-TBC (RTBC) in a conceptually different approach with the aim to unveil underlying properties of quantum disordered antiferromagnets. In Fig.~\ref{fig:Lattice}, we sketch our philosophy. Consider a generic AF Heisenberg model in the triangular lattice. For the ordered phases of the Hamiltonian, the relative orientation of the spins is fixed due to a broken symmetry, as depicted for example in the cartoon of a 2D N\'eel phase in Fig.~\ref{fig:Lattice} (top left). If the lattice is large, the bulk spins dominate over the boundary ones imposing the order expected in the thermodynamical limit independently of the chosen boundaries. However, for small lattices this is not anymore the case, and the bulk-boundary correspondence becomes much more involved. If the lattice is small, some ordered phases cannot be accommodated in the lattice. For instance, the $4\times4$ lattice is a hypercube leading to special features in its spectrum while the $4\times3$ lattice cannot accommodate the ordered N\'eel phase along its short edge. The boundaries must, thus, be properly chosen --in accordance to the lattice geometry-- to recover the underlying symmetries of the ordered phase, Fig.~\ref{fig:Lattice} (bottom left). Now, for quantum disordered phases that are not associated to a symmetry breaking, we expect the ground state of the system in the small lattice limit to be compatible with many different boundary conditions, as schematically shown in Fig.~\ref{fig:Lattice}. The lack of local symmetry if the phase is quantum disordered avoids the a priori identification of the boundaries. This feature is illustrated with the symbols "?" in Fig.~\ref{fig:Lattice} (bottom right). Nevertheless, we can count how many RTBC lead to the same ground state energy and post-select only those in order to calculate physical quantities of interest. This post-selection of the boundary conditions \cite{Yuste2017}, together with the consequences that stem from it, crucially differentiates our method, to the best of our knowledge, from any other method based on twisted or random boundary conditions.

Specifically, for 2D spin 1/2 AF Heisenberg models, the spins lay in the XY plane and TBC correspond to adding a phase in the spins $i,j$ interacting through the boundaries:
\begin{align}
\label{Eq:RBC}
S^+_{i}S^-_{j}\rightarrow S^+_{i}S^-_{j}e^{-i\phi},\\
S^-_{i}S^+_{j}\rightarrow S^-_{i}S^+_{j}e^{+i\phi}.
\end{align}
To twist the lattice simultaneously in two directions requires two different phases $\phi_1$ ($\phi_2$), for left-right (top-bottom) boundaries, as depicted in the bottom panels of Fig.~\ref{fig:Lattice}. The spins of the lattice laying at both boundaries acquire a phase $\phi=\phi_1+\phi_2$. Notice that conventional PBC favor order commensurate with the lattice dimensions, $N=L\times W$, since in the reciprocal lattice, momentum is selected at $k_1=2\pi n_1/L$ and $k_2={2\pi n_2}/W$ for $n_i\in\mathbb{N}$. In contrast, TBC allow to test all possible momenta in the first Brillouin zone \cite{Poiblanc91,Gros92,Thesberg2014}
\begin{eqnarray}
k_{1}&=&\dfrac{2\pi n_1}{L}\pm \dfrac{\phi_1}{L},\cr
k_{2}&=&\dfrac{2\pi n_2}{W}\pm \dfrac{\phi_2}{W}.
\end{eqnarray}

Let us briefly review our approach \cite{Yuste2017}. First, we fix the lattice size $N$, and its geometry. Here, we use $N=4\times 3$ or $N=4\times 4$, but to ensure convergence, some of the results are also calculated for $N=6\times 4$ and $4\times 6$ . Then, we generate a set $p$ of two randomly chosen phases, $\{\phi_{1},\phi_{2}\}_p$, with $\phi_i\in[0, 2\pi)$ and $p=1,2,\dots,200$. For each configuration, we diagonalize the Hamiltonian, generating a ground state $\ket{\psi_p}$ with energy $E_p$, and denote by $\ket{\psi_0}$ the ground state with the lowest energy, $E_{0}$. We post-select those configurations whose ground state energy fulfills: $\epsilon_p=(E_{p}-E_{0})/|E_{0}|<\alpha$.
The election of the energy bias, $\alpha$, is somehow arbitrary as it depends on the lattice size and the ratio between bulk and boundary interactions. Nevertheless, our  results are independent of it if the set $p$ is sufficiently large. Notice, however, that for small lattices the bias cannot be vanishingly small. Note also that, since the post-selection implies that several different twisted boundaries are simultaneously used to describe the same Hamiltonian parameters, our method cannot be interpreted as the insertion of an external magnetic flux in a lattice with periodic boundary conditions. 

Consequently, one relevant figure of merit is the number of configurations, $N_c$, laying in the interval $0\leq\epsilon_p<\alpha$.  Typically, we choose $\alpha=0.01$, meaning that only configurations whose ground state energies are less than a $1\%$ higher than $E_{0}$ are retained. For ordered phases, just very few random TBC accommodate the symmetry of the phase and the ones which do not, correspond to large $E_p$ and are automatically discarded in our approach. In contrast, we find regions in the Hamiltonian parameters where $N_c$ increases dramatically. The corresponding ground states, $\ket{\psi_p}$, strongly differ one from each other, as observed by computing the overlap $O_p=|\langle{\psi_p}\ket{\psi_0}|$. Finally, as it is standard in disordered systems, we calculate the quantities of interest for each post-selected configuration and perform afterwards the corresponding average, which we denote by $\langle...\rangle_d$. The average washes out some of the spurious symmetries introduced by TBC. In ED, one quantity which can be easily obtained is the static spin structure factor
\begin{equation}
\label{Eq:Sk}
S(\vec{k})=\frac{1}{N}\sum_{i,j} e^{-i\vec{k}\cdot\left(\vec{r}_i-\vec{r}_j\right)}\langle{S}_i {S}_j \rangle,
\end{equation}
where the expectation value is taken over the corresponding ground state $\ket{\psi_p}$. From the spin structure factor, one can extract the following order parameter
\begin{align}
\label{Eq:M}
M=\sqrt{S(\vec{Q}_{\rm{max}})/N} \;,
\end{align}
where $\vec{Q}_{max}$ are the k-vectors corresponding to the maxima of the spin structure factor in the first Brillouin zone. This parameter signals long range order (LRO) and, therefore, the presence of a quantum disordered phase must be accompanied by a decrease of LRO. Regarding entanglement, it is well known that local entanglement measurements cannot detect QSL, but they help to identify the underlying ordering of the phases. 
Aside from the topological entanglement entropy, topological properties can also be detected through the entanglement spectrum~\cite{PhysRevLett.109.237208,1742-5468-2014-10-P10008,PhysRevX.4.041028}. However, in 2D systems the entanglement spectrum depends explicitly on the particular chosen partition and it becomes cumbersome to extract topological properties with this method \cite{1742-5468-2014-10-P10008}. To release such constrain, we analyze here both, the concurrence and entanglement entropy. The concurrence is a measure of entanglement between any two spins $i,j$ of the ground state described by its reduced density matrix $\rho_{ij}=\tr_{k\ne i,j}(\ket{\psi_p}\bra{\psi_p})$, which can be easily computed in ED methods \cite{Wooters97}. For local Hamiltonians, the concurrence cannot capture long range entanglement \cite{De_Chiara_2018}. To go beyond short range entanglement, one can use also geometrical entanglement. It "measures" the distance of a state to its closest separable one
\begin{equation}
\Lambda_{\rm max,p} = {\max_{\ket{\phi_{\rm prod}}} |\langle{\psi}_p|{\phi_{\rm prod}}\rangle|}
\end{equation}
where $\ket{\phi_{\rm prod}}=\otimes_{i=1}^N \ket{\phi_i}$, and we maximize over the set of all separable (non-entangled) states.
The larger $\Lambda_{\rm max,p}$, the lower entanglement of $\ket{\psi_p}$, since it is closer to a product state. It makes sense to define the geometric entanglement~\cite{Wei2003} as:
\begin{equation}
\label{Eq:geometrical}
E_G= 1-\langle\Lambda_{\rm max,p}\rangle_{d},
\end{equation}
where the average over all post-selected configurations has been used. Clearly, the geometric entanglement goes beyond bipartite entanglement, and provides a measure of the amount of entanglement encoded in the state. Finally, we analyze the behavior of the ground state of the system in the small lattice under an external magnetic field. This tool has been used to elucidate the presence of topological degeneracy \cite{Thouless82, Neupert2011,Nielsen2013}.
%
\begin{figure*}
\includegraphics[width=1.4\columnwidth]{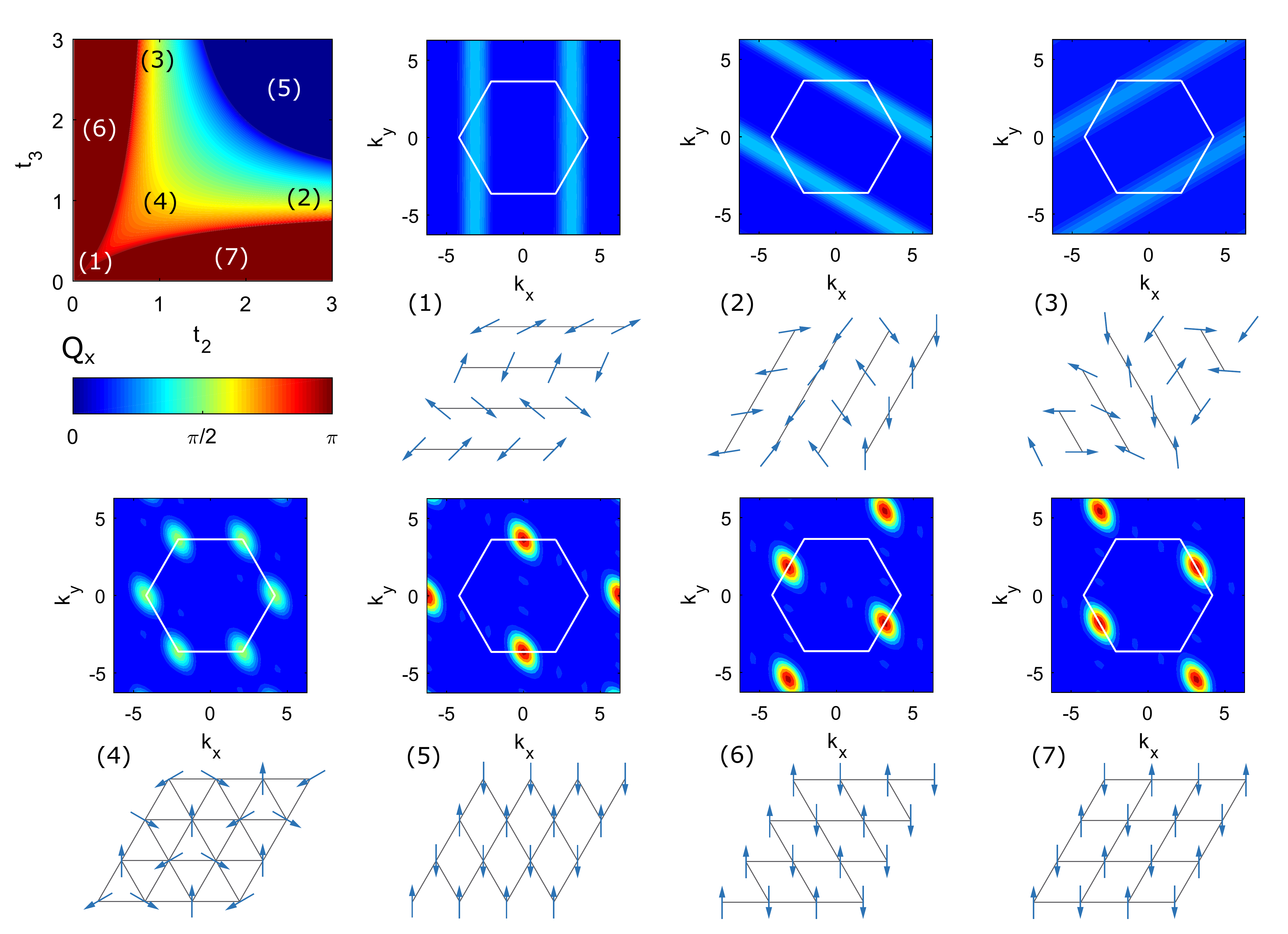}
\caption{Classical phase diagram for the SCATL for both XY ($\lambda=0$) and Heisenberg ($\lambda=1$)  interactions, obtained by plotting $Q_x$ in Eq.~\eqref{Eq:Qx_clas} as a function of the anisotropy (top left). The other panels show the spin structure factor and a sketch of the spin order for each classical phase.}
\label{fig:classical_pd}
\end{figure*} 
\section{Spatially Completely Anisotropic Triangular Lattice (SCATL)}
\label{Sec:SCATL}

Our staring point is the AF Heisenberg spin 1/2 model in a triangular lattice, whose Hamiltonian reads:
\begin{align}
\label{eq:Hamiltonian}
{H}=\sum_{<i,j>}t_{ij}\,({S}^x_i{S}^x_j+{S}^y_i{S}^y_j+\lambda\,{S}^z_i{S}^z_j),
\end{align}
where ${S}^\alpha_i$ are the spin 1/2 Pauli matrices for site $i$, the sum runs over all NN pairs, and the notation $t_{ij}>0$ denotes the coupling constants (i.e., tunneling in the corresponding Bose Hubbard model). We restrict ourselves to the cases $\lambda=0$ ($\lambda=1$), which correspond to XY (Heisenberg) interactions. 
The anisotropy of the model is given by the different interaction strengths  $(t_1,t_2,t_3)$ along the lattice directions (see Fig.~\ref{fig:Lattice}, bottom). Without loosing generality, we consider $t_1=1$ and leave as free parameters $t_2$ and $t_3$. The case $t_2=t_3$ has been extensively studied~\cite{Schmied2008, Yunoki2006, Yuste2017, Hauke2010}. For the sake of completeness, it is instructive to reproduce first its classical phase diagram. The reader familiar with it can skip this part.

\noindent {\it {Classical Phase Diagram.}}
\begin{figure*}
	\includegraphics[width=1.4\columnwidth]{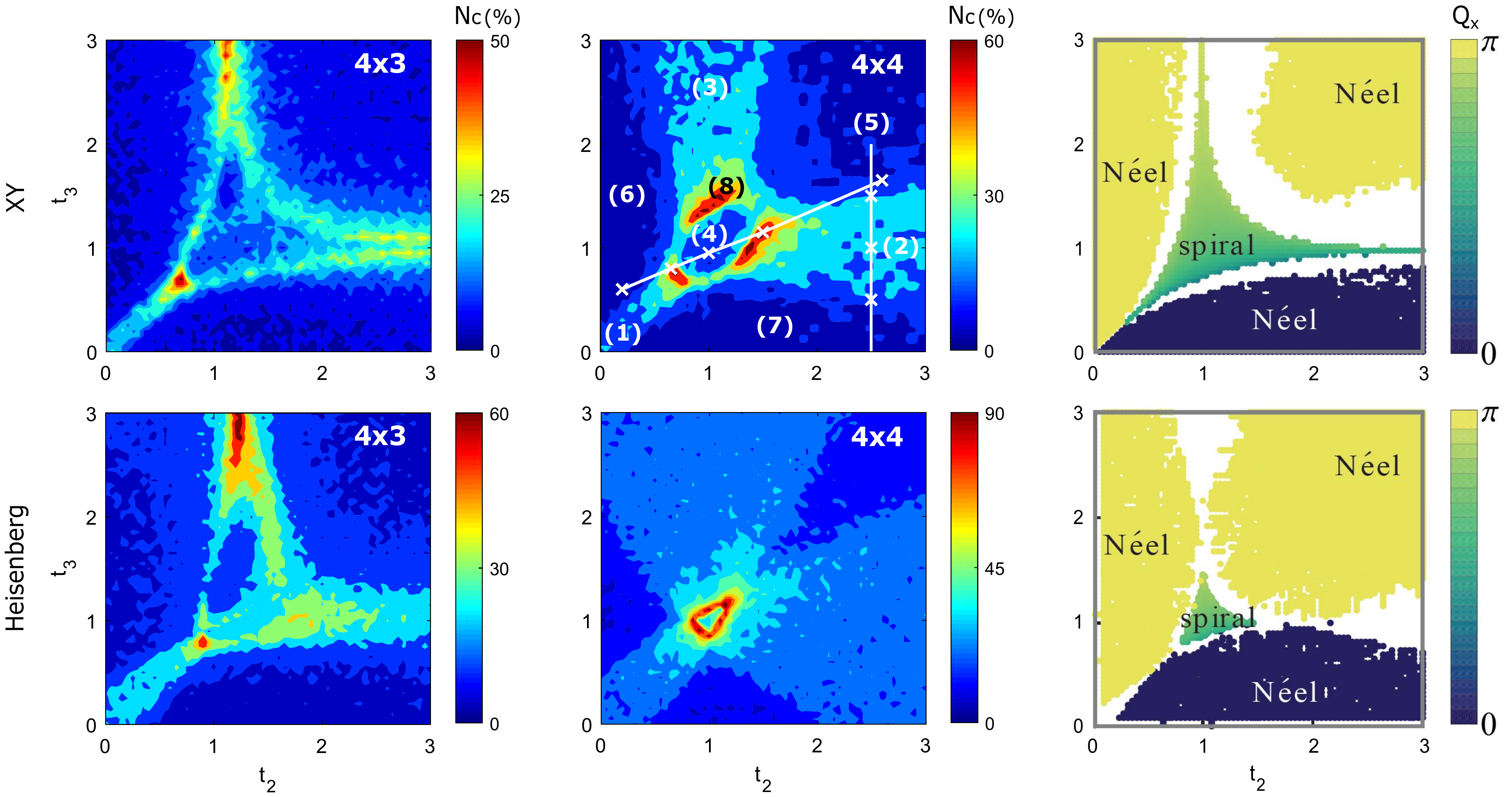}
	\caption{Quantum phase diagram of the SCATL model as obtained through the number of quasi degenerate configurations, $N_c$, for a $4\times3$ and $4\times 4$ lattice. Upper (lower) panels, XY (Heisenberg) interactions. In the $4\times 4$ XY diagram, the different quantum phases are labeled $1$ to $8$ (see text for details). The right panels correspond to the quantum phase diagram obtained with MSWT from Ref.~\cite{Hauke2013A}, where the white regions correspond to the breakdown of the MSWT calculations and are indistinguishably associated to gapless or gapped QSL.}
	\label{fig:NC}
\end{figure*}
 The classical phase diagram provides an estimate of the location and nature of the ordered phases. Order is signaled by the points in the reciprocal space that maximize correlations or, equivalently, the ones that minimize the Hamiltonian energy. The classical ordering vector, $\vec{Q}^{\rm cl}$, is obtained replacing the spin operators in Eq.(\ref{eq:Hamiltonian}) by a classical rotor laying in the XY plane,  $S_i=S\cdot\left(\cos\left(\vec{Q}^{\rm cl}\cdot\vec{r}_i\right),\sin\left(\vec{Q}^{\rm cl}\cdot\vec{r}_i\right)\right)$, up to a global phase. Energy minimization yields a region in the phase diagram with continuously varying ordering vector, described by the following equations:
\begin{eqnarray}
\label{Eq:Qx_clas}
Q_x^{\rm cl}&=&\pm\arccos\left[\frac{t_2t_3}{2}-\frac{t_2^2+t_3^2}{2t_2t_3}\right] \quad {\rm if}\quad \left|\frac{t_2t_3}{2}-\frac{t_2^2+t_3^2}{2t_2t_3}\right| \leq 1 \nonumber\\
Q_y^{\rm cl}&=&\pm\frac{2}{\sqrt{3}}\arccos\left[\mp\left(\frac{t_2+t_3}{2t_2t_3}\right)\sqrt{t_2t_3+2-\frac{t_2^2+t_3^2}{t_2t_3}}\right]\,,
\end{eqnarray}
where the argument of $Q_y^{\rm cl}$ is negative if the corresponding $Q_x^{\rm cl}$ satisfies $|Q_x^{\rm cl}|\leq \pi$, and positive otherwise. 
\begin{figure}
	\includegraphics[width=0.8\columnwidth]{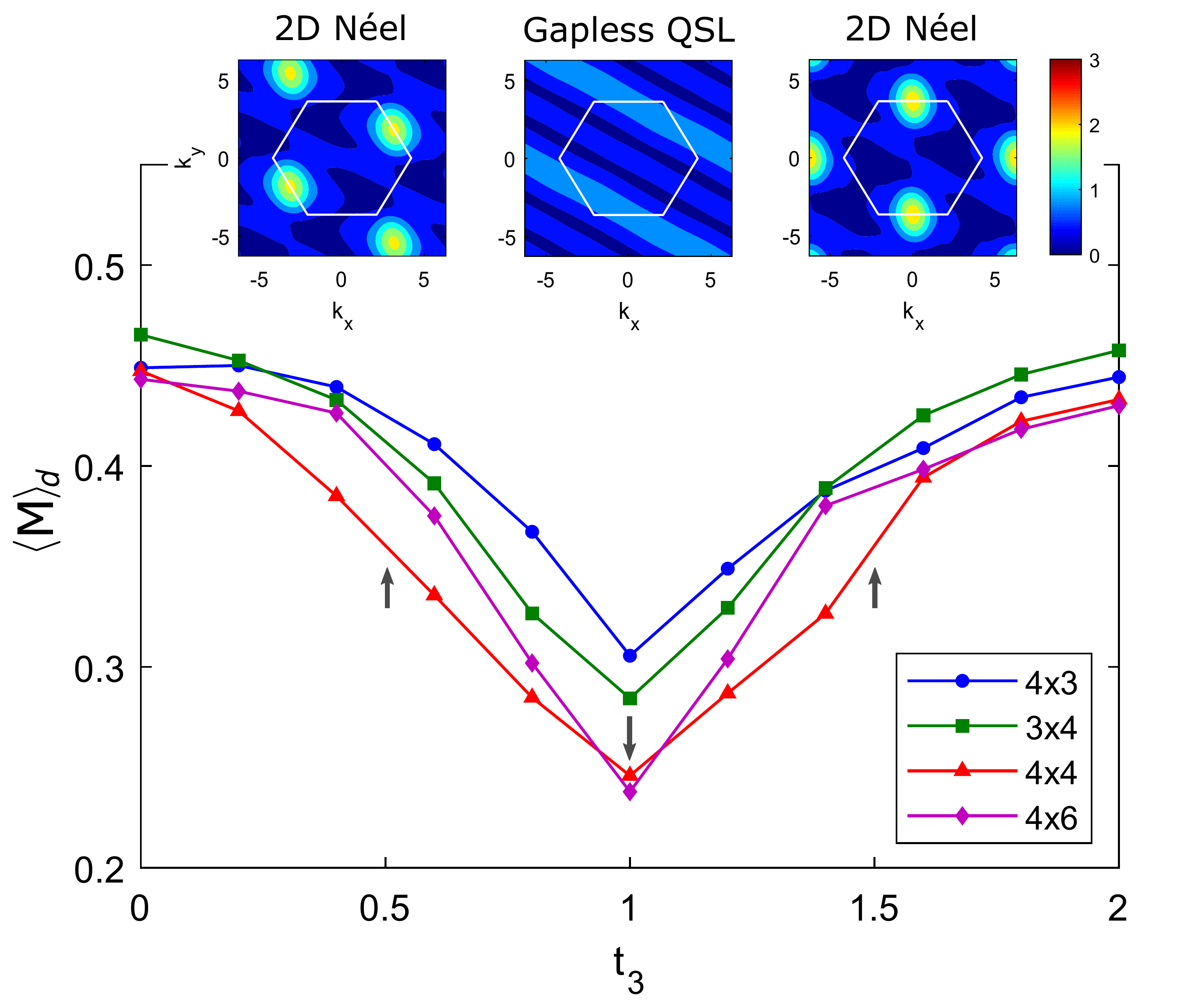}
	\caption{Study of the SCATL quantum phase diagram in the region between two 2D N\'eel phases. The averaged order parameter, as defined in Eq.(\ref{Eq:M}), is plotted along the vertical white line in Fig.~\ref{fig:NC} for different lattices sizes. The three arrows indicate the points where the three representative spin structure factors are plotted.}
	\label{fig:Neel_Gapless}
\end{figure}
The classical phase diagram is depicted in Fig.~\ref{fig:classical_pd}, together with the representative spin structure factor of each phase. First, we describe the 1D lattice limit corresponding to \text{(1)}\; $t_2=t_3=0\;$; \text{(2)} $\;t_2\to \infty, t_3=1$; and  \text{(3)}$\;t_3\to \infty, t_2=1$ as shown in  Fig.~\ref{fig:classical_pd}. For these cases, the lattice becomes a system of uncorrelated chains and the corresponding phases are 1D N\'eel ordered along the dominant lattice coupling and uncorrelated along the other two. This is clearly shown in the corresponding spin structure factors. 
At the isotropic point, $t_2=t_3=1$, indicated by \text{(4)} in Fig.~\ref{fig:classical_pd}, the system has spiral order (N\'eel 120$^{\rm{o}}$) with maxima in the structure factor at all the vertices of the reciprocal lattice cell.  
This phase extends as an incommensurate spiral phase merging smoothly with the classical 2D N\'eel phases corresponding to $t_i=t_j>>t_k$, and the  lattice deforms into diamond lattices along the two dominant directions, indicated in Fig.~\ref{fig:classical_pd} by \text{(5,6,7)}. This completes the classical phase diagram. Finally, we also add a symbolic sketch of the spin orientations for each phase.
\begin{figure*}
	\includegraphics[width=1.4\columnwidth]{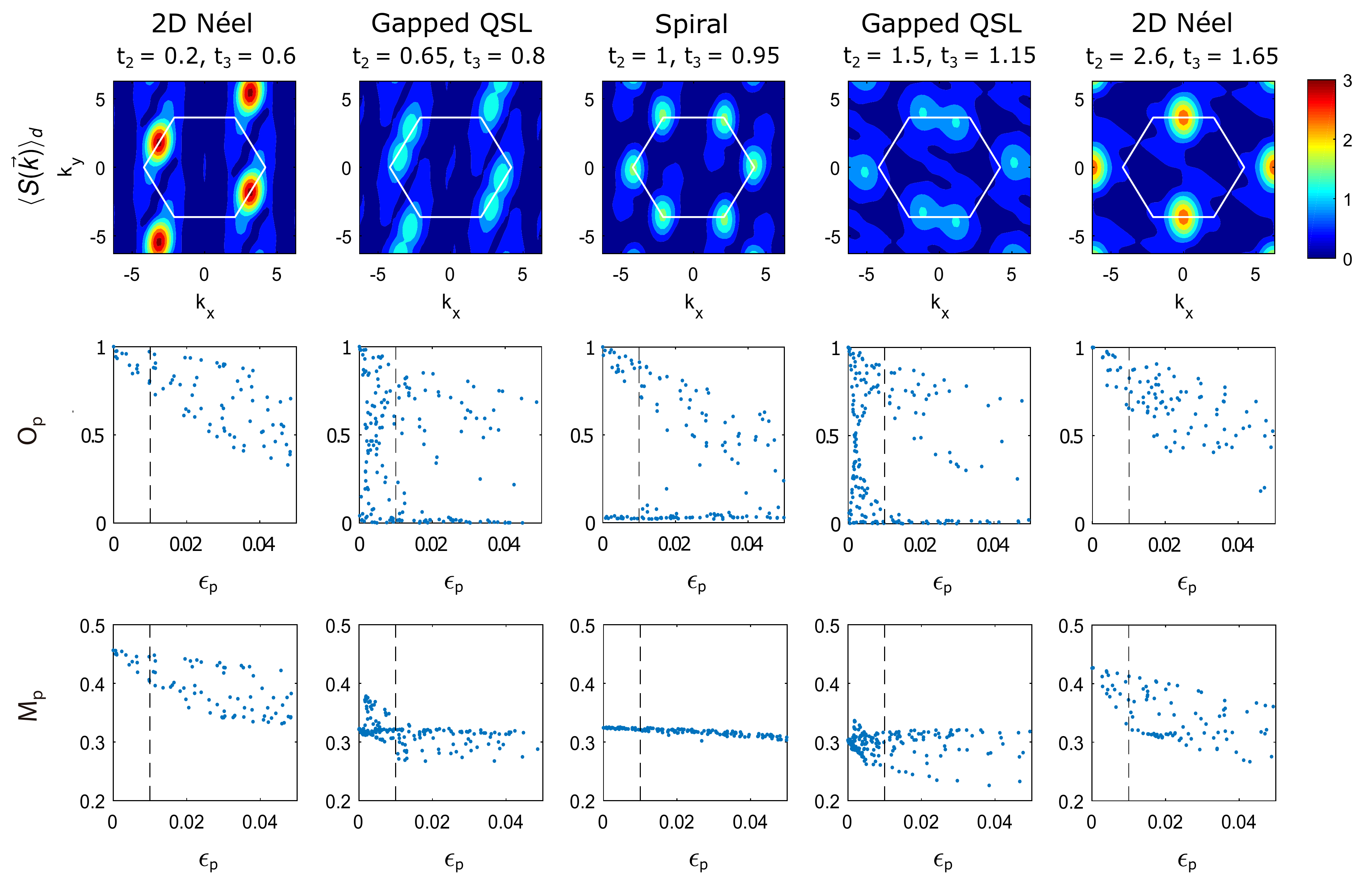}
	\caption{Quantum phase diagram along the diagonal white line in Fig.~\ref{fig:NC} for the $4\times4$ lattice with XY interactions. Top: averaged structure factor $\langle S(\vec{k})\rangle_d$. Center: overlap, $O_p$, versus relative energy, $\epsilon_p$. Bottom: order parameter, $M_p$ (see text). The dashed vertical lines limit the region $\epsilon_p<0.01$, where the average is done.}
	\label{fig:Overlap_SL}
\end{figure*}

\noindent{\it{Quantum phase diagram.}}
Our first results for both XY and Heisenberg interactions are summarized in the schematic phase diagram of Fig.~\ref{fig:NC} (column 1 and 2). Our figure of merit there is $N_c$, i.e., the number of configurations such that $\epsilon_p<0.01$ for a lattice of size $N=4\times3$ and $N=4\times4$. For the sake of comparison, we also plot in the last column of Fig.~\ref{fig:NC}, the quantum phase diagram obtained with MSWT from ~\cite{Hauke2013A}. The dependence of our results on the lattice size prevents a precise location of the phase boundaries, but as we shall see, it does not change their characterization.  

Let us first focus on the $N=4\times4$ lattice case (Fig.~\ref{fig:NC}, second column) for both the XY and Heisenberg models. While the figure of merit $N_c$ simply tells the number of energetically close configurations with different boundary conditions, further analysis of the quantum phase diagram demands computing the spin structure factor, the order parameter and the entanglement properties.

In accordance to the classical phase diagram, spiral ordering (labeled by (4) in Fig.~\ref{fig:NC}) occurs around the isotropic point $t_3=t_2=1$, and its extension is much reduced as compared to the classical case, in particular in the Heisenberg model. Surrounding the spiral phase, we observe a region, absent in the classical phase diagram, with a massive number of energetically compatible ground states (labeled by (8) in Fig.~\ref{fig:NC}). This is a signature of a disordered quantum phase and it is reconcilable with the conjectured gapped QSL reported in  \cite{Yunoki2006,Schmied2008,Hauke2010,Hauke2013A,Celi2016} for the isotropic line $t_2=t_3$. 
Continuously connected to this "gapped QSL" phase, there are three regions labeled by {\text(1, 2, 3)} in Fig.~\ref{fig:NC}. These regions lay between two 2D N\'eel ordered phases (\text{5,6,7}) that span around $t_i=t_j>>t_k$, and are connected to the respective classical 1D limit of uncoupled chains: $t_i\rightarrow\infty$,  $t_j=t_k$. The regions {\text(1, 2, 3)} are commonly referred in the literature as gapless QSL, and are not particularly enhanced in Fig.~\ref{fig:NC} because, in them,  the spins are ordered along the corresponding dominant direction and totally disordered along the other two. This constraint strongly restricts the number of random TBC which are quasi degenerate in energy. However, an inspection of the corresponding ground states shows that they are indeed 1D disordered quantum phases. All our results apply both to the XY and the Heisenberg model, but for the sake of concreteness, we refer from now on to the XY model.

In Fig.~\ref{fig:Neel_Gapless}, we plot the averaged order parameter $\langle M\rangle_d$ for different lattice sizes along the vertical line displayed in Fig.~\ref{fig:NC}, which goes from a 2D N\'eel state with $t_3= t_2>>t_1$ (5) to a 2D N\'eel state occurring for $t_2>>t_3$ (7). In both N\'eel phases, the value we obtain, $\langle M\rangle_d \simeq 0.44$, closely matches the value obtained in the square lattice limit, i.e., $t_i=t_j>>t_k$, with precise QMC calculations, $M=0.4373$ ~\cite{Sandvik99}. Between the two 2D N\'eel ordered phases, faithfully identified by the spin structure factor and the order parameter, there is a region with lower LRO signaled by the decrease of $\langle M\rangle_d$. The value of $\langle M\rangle_d$, although finite, clearly decreases as the lattice size increases, suggesting $M\rightarrow 0$ in the thermodynamic limit. Furthermore, the corresponding spin structure factor shows the expected pattern for 1D N\'eel order. We identify this region as a trivial gapless QSL. The same features are observed in the two other limiting cases (1) and (3).

\begin{figure}
	\includegraphics[width=0.8\columnwidth]{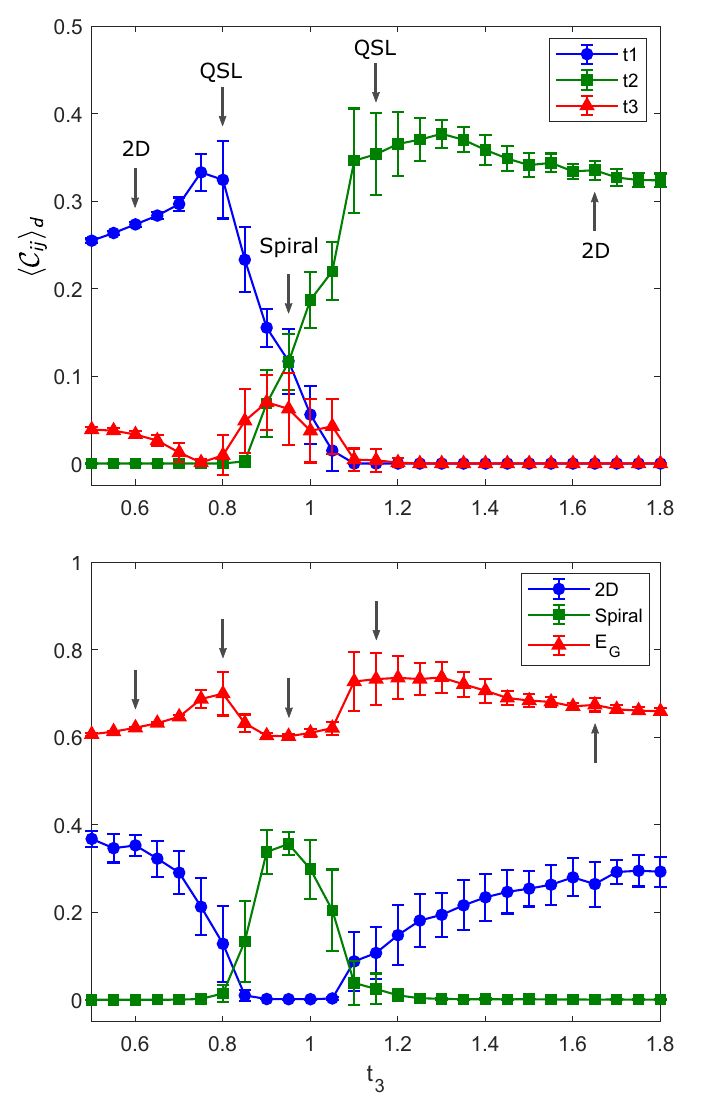}
	\caption{Top: averaged concurrence between NN sites along the three lattice directions. Bottom: geometrical entanglement, $E_G$ (Eq.~\eqref{Eq:geometrical}), and projections to separable states (see text). All quantities are averaged over the configurations with $\epsilon_{p}<0.01$, and the dispersion of the values is represented by error bars. The plotted region corresponds to the white diagonal line in Fig.~\ref{fig:NC}, where the values of $t_2$ are chosen accordingly. The arrows indicate the values used in Fig.~\ref{fig:Overlap_SL}, which are representative of each quantum phase explored.}
	\label{fig:entanglement}
\end{figure}
To further explore the nature of the truly quantum disordered phase, we restrict now our analysis to the quantum phase diagram along the diagonal line depicted in Fig.~\ref{fig:NC}, which crosses several quantum phases including the assumed gapped QSL (8). 
In the top row of Fig.~\ref{fig:Overlap_SL}, we display $\langle S(\vec{k})\rangle_d$ for some selected points along this line. Its inspection allows for an easy identification of two 2D N\'eel phases at the extremes of this quantum phase diagram. The first one exemplified at $t_2=0.2,\; t_3=0.6$, and the second one at $t_2=2.6,\; t_3=1.65$. Between them, we find the expected spiral phase at $t_2=1,\;t_3=0.95$. Finally, between the 2D N\'eel phases and the spiral one, there are two regions (circa  $t_2=0.65,\;t_3=0.8$ and $t_2=1.5,\; t_3=1.15$), whose spin structure factor does not correspond to any order. In the middle row of the same figure,  we plot the corresponding overlap  $O_p=|\bra{\psi_p}\psi_0\rangle|$ for all configurations $p$, sorted by their energy.  The energy bias for post-selection is there indicated by a dashed vertical line. For the 2D N\'eel order, $O_p$ slowly decreases as $\epsilon_p$ increases, meaning that quasi-degenerate states correspond to alike ground states. A similar behavior is observed for the spiral phase, except $O_p$ has two branches around $O_p=1$ and $0$. They correspond to the two orthogonal chiralities of the spiral ground state. In contrast, the "gapped QSL" phase shows a radically different behavior. All set of  post-selected configurations (i.e., $\epsilon_p<0.01$) might correspond to very different ground states.

Finally, in the last row of Fig.~\ref{fig:Overlap_SL}, we display the value of 
the order parameter $M_p$ (as defined in Eq.(\ref{Eq:M})) for all configurations prior to any averages. While ordered quantum phases have a very small dispersion of the order parameter, the dispersion becomes much more significant for the presumptive gapped QSL, indicating that there is not a well defined value of the order parameter in these regions.\\
We proceed by calculating the entanglement properties for the same parameters of Fig.~\ref{fig:Overlap_SL}. In Fig.~\ref{fig:entanglement} (upper panel), we show the averaged concurrence, ${\mathcal{C}}_{ij}$, between NN along the three lattice directions ($t_1,t_2,t_3$), as well as its dispersion. The vertical arrows in the figures indicate the location of the different quantum phases (2D N\'eel--QSL--spiral--QSL--2D N\'eel ) under study in Fig.~\ref{fig:Overlap_SL}.
As expected, the spiral phase has an isotropic concurrence along all directions. The concurrence also signals the two preferred directions in the 2D N\'eel phases. 

In the bottom panel of Fig.~\ref{fig:entanglement}, we display the geometrical entanglement $E_G$, together with the projection of the post-selected ground states $\ket{\psi_p}$ on a classical 2D N\'eel state and a classical spiral state ($120^{\rm{o}}$ N\'eel). As shown there, the assumed QSL phases display a larger geometric entanglement as compared to the surrounding ordered phases and a vanishing overlap with classical states.

A unique topological feature of 2D gapped quantum spin liquids is the presence of non-trivial Chern numbers or topological invariants. For many-body systems in a lattice, the Chern numbers can be straightforward computed \cite{Niu85,Fukui2005}. However, since we associate the presence of a gapped QSL with the presence of many different compatible ground states, each of them associated to different twisted boundary conditions, calculating the Chern number becomes very involved. Another characteristic feature of topological states is a ground state degeneracy that depends on the topology of the surface on which the states are defined. For a lattice with periodic boundary conditions, i.e., a torus, there exists the possibility to check such topological degeneracy by inserting an "artificial" magnetic flux perpendicular to the torus geometry that simulates the phase acquired by the atoms when they loop along the transverse direction. The topological degeneration can be understood from the similarity with the fractional quantum Hall effect (FQHE), and its mapping to the corresponding Laughlin state in the thermodynamical limit. If the corresponding Laughlin  state has filling factor $\nu=1/2$, the topological degeneracy in the thermodynamical limit will be equal to $2$ (see e.g \cite{Regnault2014}). For small lattices normally this degeneracy is not seen, but by inserting an external magnetic flux, it is possible to check if there is a flow of one ground state onto another \cite{Thouless82,Nielsen2013}. In the case of $\nu=1/2$, the level crossing will manifest for an external magnetic flux $\Phi=\pi$. Moreover, the gap to higher energy levels will remain finite for any value of inserted flux $\Phi$. Similar calculations for larger lattices (e.g. $6\times 5$) have been done for other models \cite{Nielsen2013} to show the topological character of the phase.
\begin{figure}
	\includegraphics[width=0.6\columnwidth]{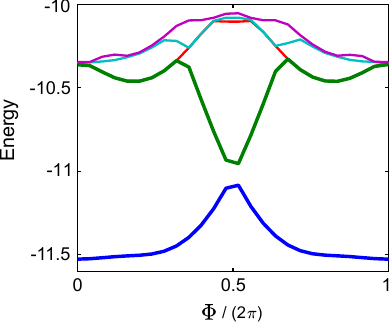}
	\caption{Energy spectrum of the five lowest states in the $S_z=0$ manifold for a $4\times4$ lattice in the putative QSL ($t_2=t_3=0.7$) using PBC ($\phi_1=\phi_2=0$) as a function of an inserted twisting phase, $\Phi$, on the boundary along the horizontal direction simulating an external magnetic field.}	\label{fluxPBC_plot}
\end{figure}
\begin{figure}
	\includegraphics[width=0.99\columnwidth]{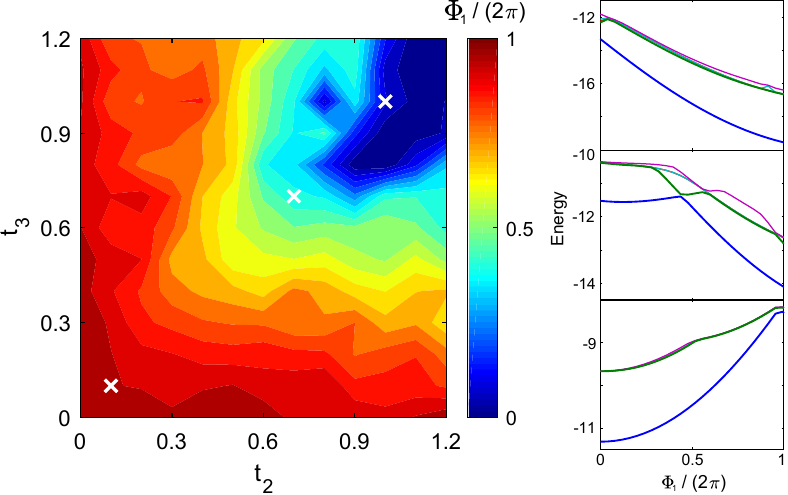}
	\caption{Left: introduced external magnetic flux at which the minimum gap between ground and first excited state is reached in the SCATL-XY model using RTBC in a $4\times4$ lattice. The flux  $\Phi_1$ is implemented by modifying the tunneling coefficients along the $x$-direction as explained in the text. The plotted values correspond to the average over all the compatible configurations $N_c$. Right: energy spectrum of the five lowest states in the $S_z=0$ manifold as a function of the inserted flux $\Phi_{1}$  for one single configuration of our RTBC ($\phi_{1},\phi_{2}$). The plots correspond to the points indicated with white crosses on the left panel. From bottom to top, (i) 1D N\'eel ($t_2=t_3=0.1$), (ii) expected gapped QSL ($t_2=t_3=0.7$), and (iii) spiral phase ($t_2=t_3=1$). Clearly, the gap is only clossing in the expected gapped QSL phase.
		}	\label{flux_plot}
\end{figure}
In this spirit, we analyze our system for a representative point of the predicted QSL phase using first PBC and twisting the boundary along the horizontal direction with a phase $\Phi \in [0,2\pi)$. In Fig.~ \ref{fluxPBC_plot}, we display the energy spectrum for the five lowest eigenstates (i.e., $S_z=0$ manifold) on the torus as a function of the inserted twisting phase $\Phi$ for a $4\times4$ lattice. The crossing of the two lowest levels, although not perfect, can be clearly appreciated. The gap to high energy levels remains finite for any value of the inserted flux. 

Now, for all the compatible configurations $N_c$ given by our RTBC (i.e., all compatible $\phi_1,\phi_2$ leading to almost degenerate ground state energy), we simulate the insertion of the external magnetic flux by modifying the tunneling couplings along the $x$-direction as:
\begin{equation}
t_1\longrightarrow t_1e^{i\Phi_1/L} \;, \label{Eq:introduceflux}\\
\end{equation}
with $L$ the number of spins along the $x$-direction, and $\Phi_1\in [0,2\pi)$. We compute the energy spectrum for each configuration as a function of the phase, $\Phi_1$, and extract the flux for which the gap between the first and second eigenvalues is minimum. Finally, we average the results over all considered configurations $N_c$. Our results for the SCATL-XY model are displayed in Fig.~\ref{flux_plot} for a significant part of phase diagram using a $4\times 4$ lattice. 

Although the closing of the gap is not complete (as it happens already for the PBC of Fig.~\ref{fluxPBC_plot}),  it is interesting to notice that for the phase diagram regions compatible with 2D- N\'eel phases or 1D-gapless QSL, the average gap closes trivially for a flux of $\Phi_1\simeq 2\pi$. In these regions, the effect of the anisotropy reduces the triangular lattice into a set of disconnected one-dimensional chains or into a "squared" lattice, suppressing in this way frustration. At $\Phi_1=2\pi$, there are two compatible ground states related by a flip of all spins. For the putative gapped QSL, the gap becomes minimal for a flux $\Phi_1=\pi$. As it can be seen in Fig.~\ref{flux_plot}, the closing of the gap for a flux of $\Phi_1\simeq\pi$ occurs for all the phase diagram surrounding the spiral phase. Finally, in the spiral phase, the gap does not close for any value of the external flux, indicated in the figure by $\Phi_1=0$. As we explain later, our numerical results show that the gapped QSL in the completely anisotropic Heisenberg model (SCATL) can be connected to the chiral QSL of the $J_1$-$J_2$ model and, thus, they share the same degeneracy.\\

\begin{figure*}
\includegraphics[width=1.4\columnwidth]{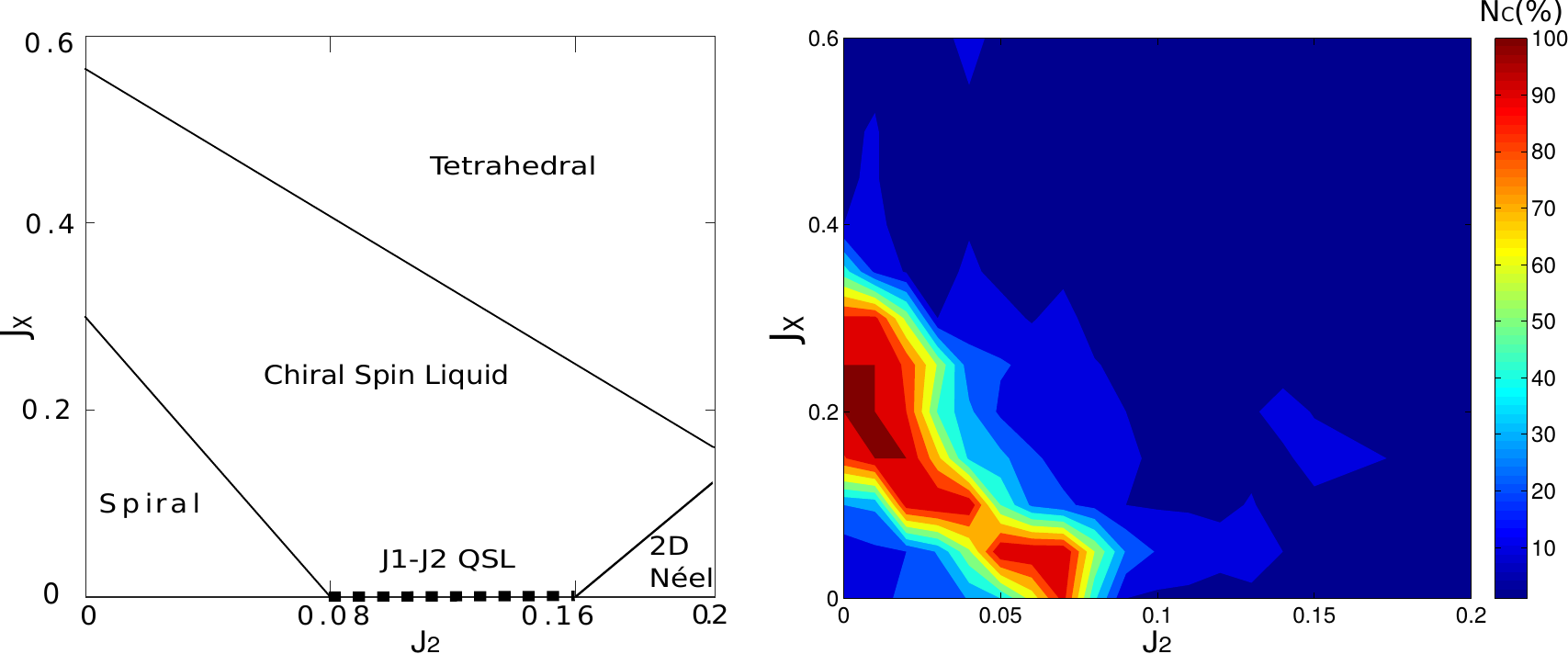}
\caption{Left: sketched quantum phase diagram of the $J_1-J_2$ model with chiral interactions obtained from Refs.\cite{Wietek2017,Gong2017}. Right: $N_c$ for the same model using a $4\times4$ lattice. The area with large $N_c$ is a signature of putative gapped QSL phase.}
\label{fig:J1J2_pd}
\end{figure*}

\section{J1-J2 model}
\label{Sec:J1J2}

In this section, we extend our work to the AF Heisenberg model with nearest (NN) and next-nearest-neighbors (NNN) interactions.
\begin{align}
\label{Eq:Ham_J1J2}
H_{J_1-J_2}=J_1\sum_{\langle i,j\rangle}\vec{S}_i\cdot\vec{S}_j+J_2\sum_{\langle \langle i,j\rangle \rangle}\vec{S}_i\cdot\vec{S}_j,
\end{align}
where we fix $J_1=1$ leaving $J_2$ as the free parameter, and the sums run over all NN and NNN pairs respectively.\\

Before proceeding further, let us mention that finite size effects are now further enhanced by the presence of NNN terms. However, in consonance with Sec.\ref{Sec:SCATL}, our aim here is to find the signatures of the ground states which are compatible with QSL rather than to provide the precise location of the quantum phase boundaries. It is important also to stress that implementing random TBC for a Hamiltonian hosting both NNN and chiral interactions, as we will later introduce, demands some subtleties which are explained in the Appendix.\\

\noindent{\it{Classical phase diagram.}}

The classical phase diagram of this system is well known~\cite{Jolicoeur90}. For $J_2<1/8$, there is a three-sublattice 120 N\'eel ordered ground state (spiral order). For $1/8<J_2<1$ the classical phase diagram is degenerate with the three different collinear 2D N\'eel order 
and a tetrahedral non-coplanar state. However, an order-by-disorder mechanism selects the 2D N\'eel order when quantum fluctuations are taken into account \cite{Jolicoeur90,Lecheminant95}. For $J_2>1$, there is non-commensurate spiral order. \\

\noindent{\it{Quantum phase diagram.}}

Recent studies have analyzed the quantum phase diagram of the model with special attention to the surroundings of the classical phase transition point at $J_2=1/8$ with 2D DMRG \cite{Zhu2015,Hu2015}, variational quantum Monte Carlo \cite{Iqbal2016}, exact diagonalization \cite{Wietek2017} and Schwinger-boson mean-field \cite{Bauer2017}. A consensus has been reached in identifying a QSL phase for $0.08\lesssim J_2\lesssim 0.15$. The nature of this phase, though, is still under debate. To shed more light in the issue, an extra chiral term in the Hamiltonian has been proposed \cite{Hu2016, Gong2017, Wietek2017,Saadatmand2017,Hickey2017},
\begin{align}
\label{Eq:Ham_J1J2C}
H_{\chi}=H_{J_1-J_2}+J_{\chi}\sum_{i,j,k\in{\bigtriangleup}}\vec{S}_i\cdot(\vec{S}_j\times\vec{S}_k),
\end{align}
where the sum runs over all the up and down triangles of the lattice clock-wisely. 

In Fig.~\ref{fig:J1J2_pd} (left panel), we show a sketch of the quantum phase diagram taken from Refs.~\cite{Hu2016, Gong2017, Wietek2017}. For $J_\chi=0$, we recover the $J_1-J_2$ model.
As $J_\chi$ is turned on, there is a phase transition from the QSL under debate into a chiral spin liquid (CSL), which lies between the ordered spiral, the 2D N\'eel collinear, and the tetrahedral phase.
In Fig.~\ref{fig:J1J2_pd} (right panel), we show our schematic quantum phase diagram obtained by counting the number of post-selected configurations, $N_c$, for $\epsilon_p<0.005$, as a function of the parameters of the model, $J_2$ and $J_{\chi}$, for a lattice of just $N=4\times 4$ spins. For this model, in contrast with the analysis of previous models (Sec. \ref{Sec:SCATL}), we choose a smaller energy bias, $\epsilon_p$, for post-selection of quasi-degenerate states because the number of inner bonds is much increased as compared to the Heisenberg model. 
For $J_{\chi}=0$, we observe a region with a large number of quasi degenerate ground states that extends approximately about $0.05\lesssim J_2\lesssim 0.10$. As $J_{\chi}$ increases, this region is continuously enlarged and at $J_2=0$, it expands approximately between  $0.10\lesssim J_{\chi}\lesssim 0.4$. It is interesting to compare both figures. Although the boundaries we obtain are clearly different from those sketched in Fig.~\ref{fig:J1J2_pd} (left panel), our results show a large increase of compatible configurations in a region reconcilable with the location of both the CSL present in the model described above (Eq.(\ref{Eq:Ham_J1J2C})) and the QSL of the $J_1-J_2$ model (Eq.(\ref{Eq:Ham_J1J2})). 
 
\begin{figure}
	\includegraphics[width=1\columnwidth]{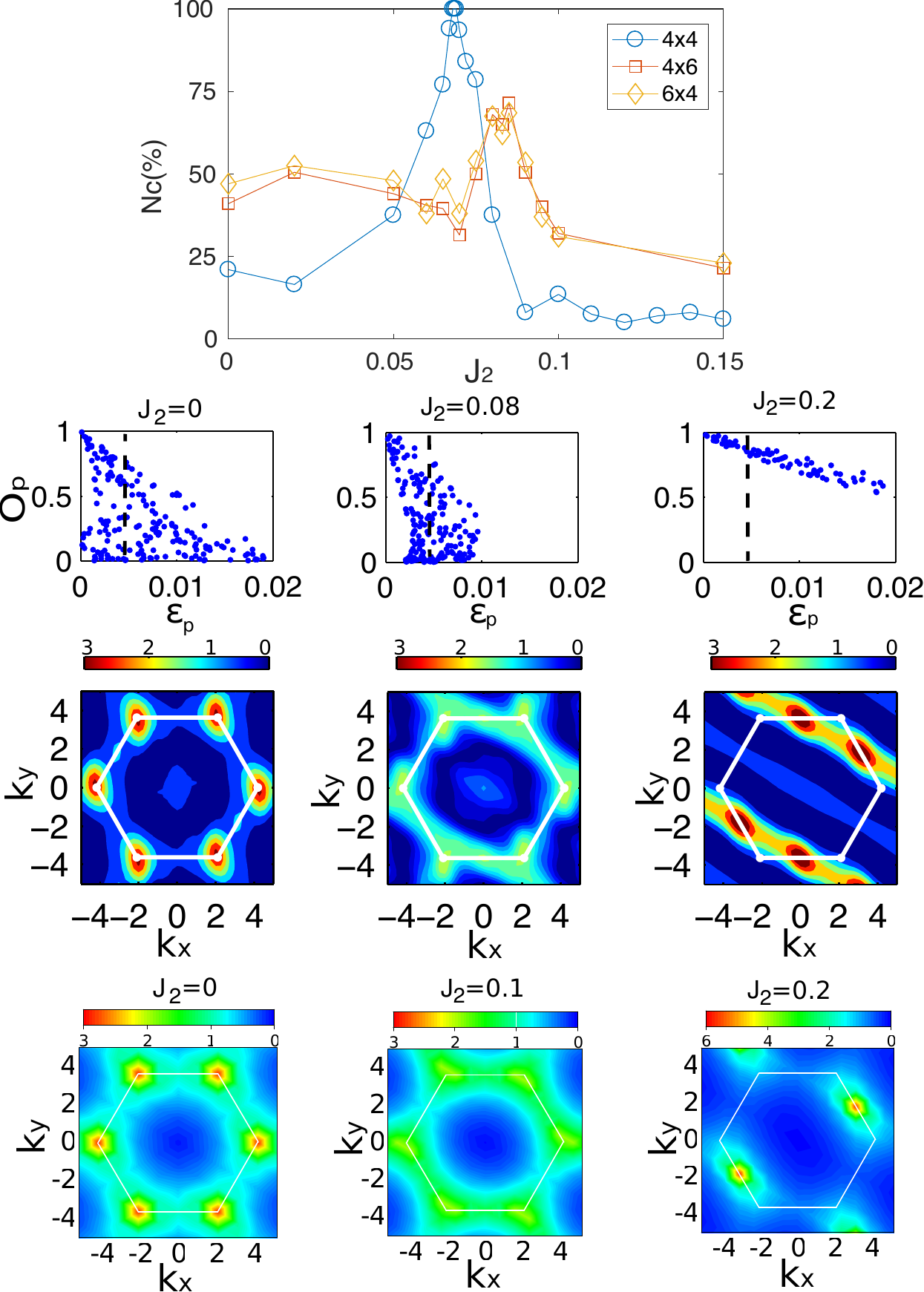}
	\caption{$J_1-J_2$ model without chiral interactions ($J_\chi=0$). Upper panel: number of configurations, $N_c$, with $\epsilon_p<0.05$ for different lattice sizes and geometries. First row: $O_p$ and energy bias $\epsilon_p$ for a $6\times 4$ lattice. 
	Second row: our average spin structure factors,  $\langle S(\vec{k})\rangle_d$. Third row: $S(\vec{k})$ obtained with 2D- DMRG taken from Ref.\cite{Hu2015}.}
	\label{fig:J1J2}
\end{figure}

Finite size effects can be spotted by calculating the quantum phase diagram in larger lattices. In Fig.~\ref{fig:J1J2} (top),  we display $N_c$ as function of $J_2$ ($J_{\chi}=0$) for different lattice sizes and geometries; $N=4\times 4$, $4\times 6$, $6\times 4$. As expected, by increasing the lattice size, the location of the maximum of $N_c$ shifts to larger values of $J_2$, in accordance to the quantum phase diagram of the system. To deepen further in the nature of the possible phases observed in Fig. {\ref{fig:J1J2_pd}} (right panel), we explore other physical quantities, like the averaged spin structure factor, to determine the corresponding orders for a lattice of $N=6\times4$. Our results are depicted in Fig.~\ref{fig:J1J2} and agree quite closely with the expected orders. For $0\leq J_2\lesssim 0.05$, spiral order is dominant. As $J_2$ further increases, there is a region with large number of random configurations, $N_c$, which lead to a ground state energy, $E_p$, quasi degenerate with the smallest one, $E_0$. These configurations correspond to different ground states, as demonstrated by all possible values the overlap $O_p$ takes.  In this region, the average structure factor, $\langle S(\vec{k}) \rangle_d$, is blurred, showing that there are no clear preferable $k$-vectors. This indicates disorder and, consequently, a decrease of LRO. Again, it is instructive to compare our results with the results of the quantum phase diagram obtained with more sophisticated methods for larger lattices. In the bottom row of Fig.~\ref{fig:J1J2}, we attach for comparison $S(\vec{k})$ obtained with 2D DMRG from Ref.\cite{Hu2015}. For the values where the putative QSL is predicted, both  $S(\vec{k})$ obtained from the 2D DMRG simulations and our $\langle S(\vec{k})\rangle_d$ are impressively similar. For $J_2=0.2$, the 2D DMRG shows collinear order corresponding to a 2D N\'eel order along two lattice directions (see  Fig.~\ref{fig:classical_pd}), while our results show a superposition of two of the 2D N\'eel collinear orders. This is not relevant, as all collinear orders are degenerate and of course any superposition of them as well. 
Finally, let us remark that in the same spirit, we have also analyzed the nature of the quantum phases that appear  when the chiral term is included for a lattice of $N=4\times4$. The results in this case suffer from strong finite size effects, but ordered phases can be easily identified by $\langle S(\vec{k})\rangle_d$.
\begin{figure}
	\centering
	\includegraphics[width=0.8\columnwidth]{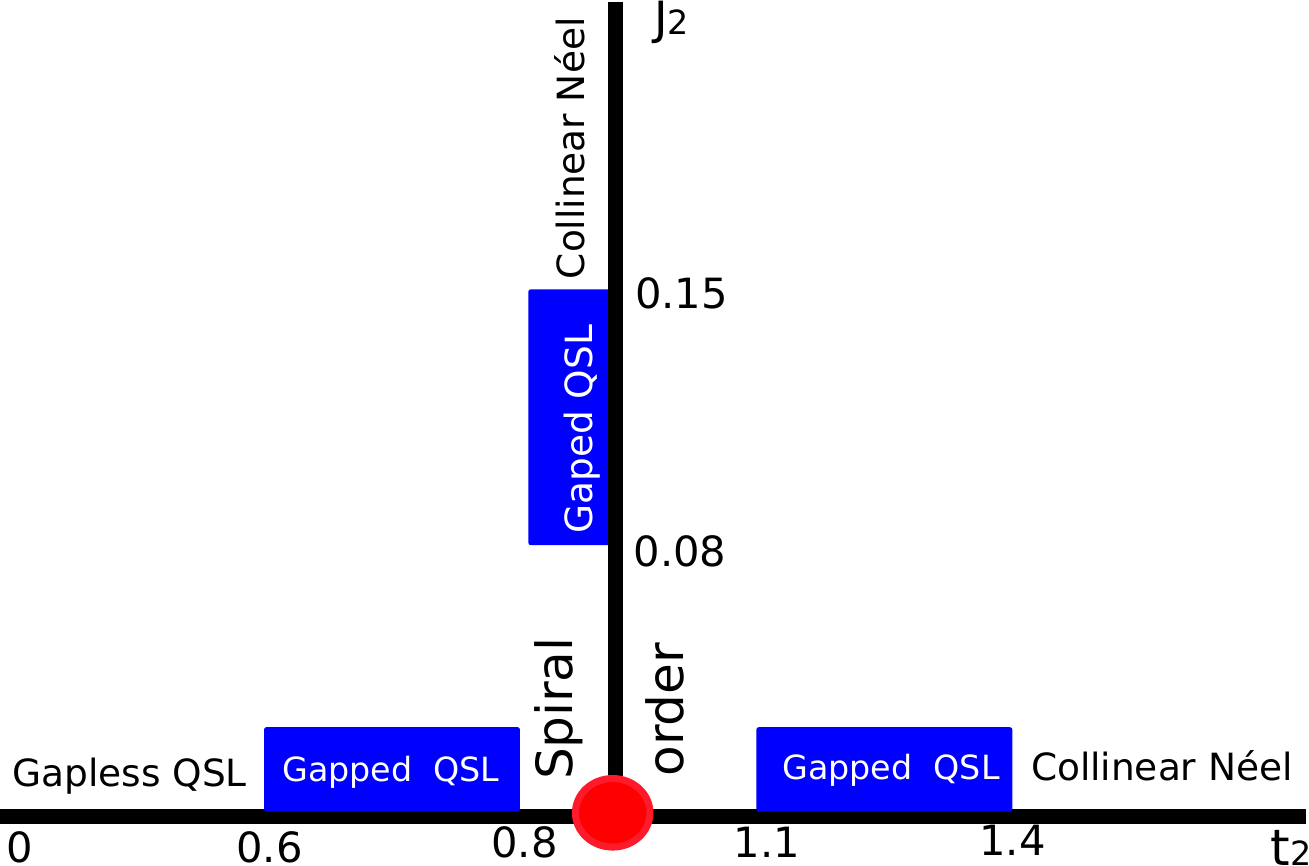}
	\caption{Sketch of the phase diagram of the hybrid anisotropic $J_1$-$J_2$ model in the triangular lattice introduced in the text.}
	\label{fig:aJ1J2}
\end{figure}
\begin{figure}
	\centering
	\includegraphics[width=0.8\columnwidth]{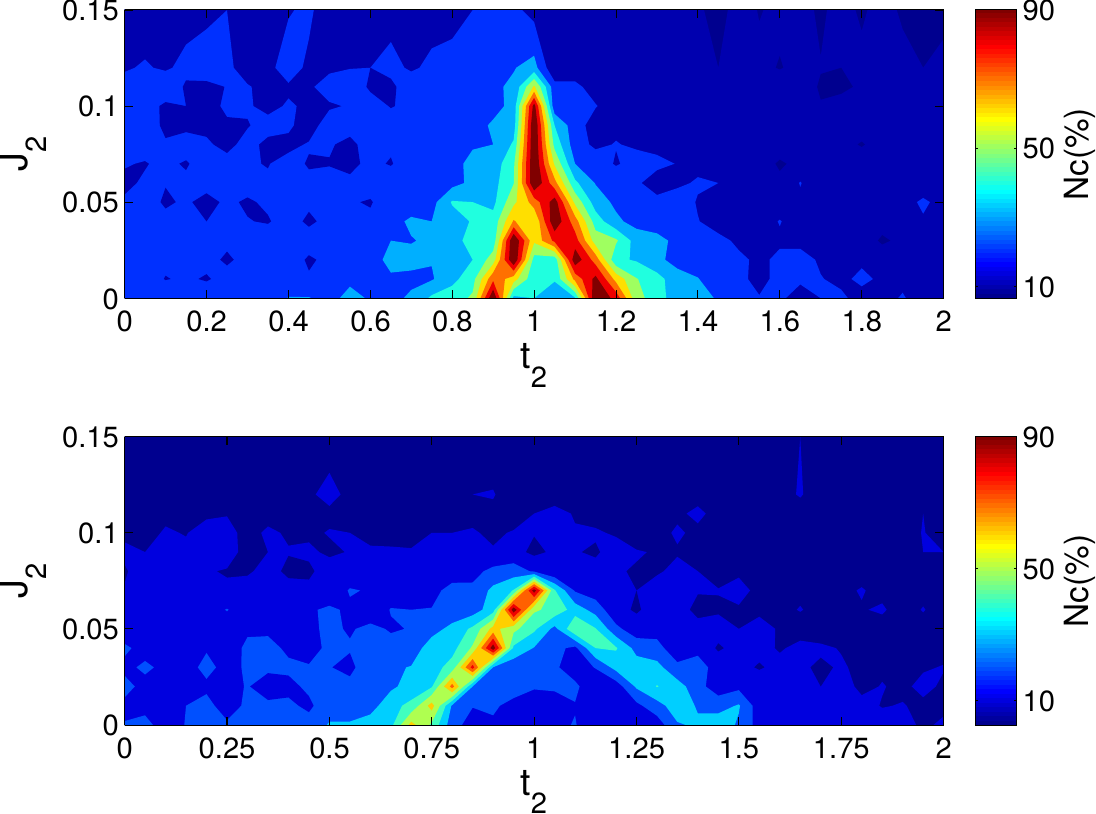}
	\caption{Phase diagram of the anisotropic $J_1$-$J_2$ model (see Eq.(\ref{Eq:aJ1J2})) obtained with random twisted boundary conditions for a plaquette $4\times 4$. Our figure of merit is $N_c$, the number of configurations with ground state energy compatible with the ground state. Top (bottom) panel corresponds to Heisenberg (XY)-interactions.}
	\label{fig:aJ1J2.2}
\end{figure}

\section{The anisotropic $J1-J2$ model}
\label{Sec:Hybrid}
To further gain insight in our method, we have proposed and  investigated a hybrid model: the anisotropic $J_1-J_2$ in the triangular lattice. This model should allow us to understand if the topological phases appearing in the SATL-XY model are connected with the topological phases of the $J_1-J_2$ model. The latter, as indicated in Fig.~\ref{fig:J1J2}, is connected to the topological chiral spin liquid, whose topological degeneracy is, in the thermodynamical limit, equal to two. Solving such a model with our RTBC method should allow us to indirectly determine if the gapped QSL of the anisotropic Heisenberg model has the same topological degeneracy in the thermodynamical limit than the chiral QSL of the $J_1$-$J_2$ model. 
The investigated hybrid model we propose reads:

\begin{eqnarray}
H&=&t_1\sum_{<i,j>}(S_i^xS_j^x+S_i^yS_j^y+\lambda S_i^zS_j^z)+\nonumber\\
&+&t_2\sum_{<i,j>}(S_i^xS_j^x+S_i^yS_j^y+\lambda S_i^zS_j^z)+\nonumber\\
&+&J_2\sum_{<<i,j>>}(S_i^xS_j^x+S_i^yS_j^y+\lambda S_i^zS_j^z),
\label{Eq:aJ1J2}
\end{eqnarray}
where the first two sums run over nearest neighbors with the anisotropic parameters $t_1$ and $t_2$ corresponding to interactions along the horizontal and diagonal bonds (SATL-XY), and the term proportional to $J_2$ indicates the next to nearest interactions. In the limit $J_2\rightarrow 0$, the model reduces to the anisotropic XY, while in the limit $t_2/t_1\rightarrow 1$, the model reduces to the $J_1$-$J_2$ model. The value $\lambda=1(0)$ corresponds to Heisenberg (XY) interactions.

A sketch of the quantum phase diagram of the hybrid model connecting the anisotropic Heisenberg model (horizontal phase diagram) and the $J_1$-$J_2$ (vertical phase diagram) is presented in Fig.~\ref{fig:aJ1J2}. Both models intersect at $J_2=t_2/t_1=1$, in the isotropic triangular lattice, whose quantum phase is spiral long range order (N\'eel-$120$). As before, we use our RTBC to diagonalize the Hamiltonian in Eq.(\ref{Eq:aJ1J2}) for a lattice of size $4\times 4$, and derive a quantum phase diagram using $N_c$ as a figure of merit. Our results displayed in Fig.~\ref{fig:aJ1J2.2} are clear: according to RTBC, the putative gapped QSL appearing in the anisotropic Heisenberg (XY) model are connected to the gapped QSL appearing in the $J_1$-$J_2$ model. Thus, we conjecture that the gapped quantum phases appearing in the anisotropic Heisenberg model are in one to one correspondence to the gapped QSL of $J_1$-$J_2$ and thus they should have the same topological degeneracy. We have as well calculated the phase diagram of this hybrid model by means of modified spin wave theory (MSWT). The application of MSWT in the triangular lattice is, however, far from trivial. These results will be presented elsewhere.

\section{Conclusions}
\label{Sec:conclusions}
We have presented a numerical method based on ED with engineered  boundary conditions (RTBC) to unveil the presence of quantum spin liquids in frustrated quantum systems in very small lattices. We have applied our method to several Heisenberg models in the triangular lattice and the quantum phase diagrams thus obtained are in qualitative accordance with previous results derived using QCM or 2DMRG. In order to elucidate the presence of gapped quantum spin liquids we have also shown that the ground states are topologically degenerated under the presence of an external magnetic. We have also proposed a new model, the \textit {anisotropic $J_1$-$J_2$} model in the triangular lattice, and calculated with RTBC its corresponding quantum phase diagram. Based on our calculations, we have conjectured that the gapped QSL phases appearing in the anisotropic XY model in the triangular lattice are the same as the ones appearing in $J_1$-$J_2$; a chiral quantum spin liquid. It will be very interesting to corroborate this conjecture with other methods.

Finally, it is important to signal that our method is not free from finite size effects although it strongly reduces them as compared to traditional ED methods. Indeed, the precise location of the distinct quantum phases found depends on the system size and improves as the size of the lattices increases. It remains as an open question if a finite size scaling can also be applied with our RTBC.

\textit{Acknowledgements}. We thank P. Hauke, G. Sierra and the referees for fruitful discussions and S. Gong for sharing their results with us. We acknowledge financial support from the Spanish MINECO projects FIS2016-80681-P and the Generalitat de Catalunya CIRIT (2017-SGR-1127). 

\bibliographystyle{apsrev4-1}
\bibliography{biblio_RBC}

\section*{Appendix}
\label{Sec:App}

\renewcommand\thefigure{A.\arabic{figure}}    
\setcounter{figure}{0}
\begin{figure}
	\vspace{0.5cm}
	\includegraphics[width=0.9\columnwidth]{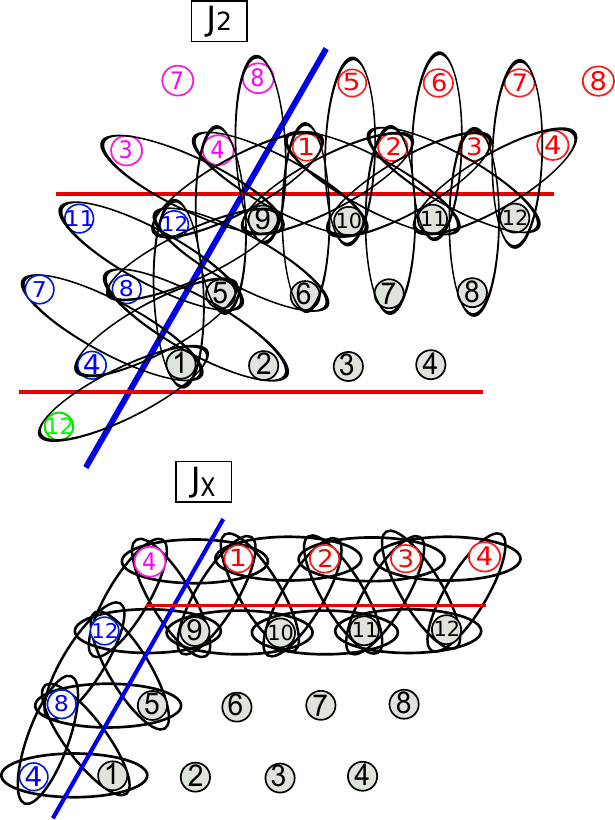}
	\caption{Scheme of twisted boundary conditions in a $4\times 3$ triangular lattice with next-nearest-neighbors interactions (top panel) and chiral interactions (bottom panel). In every interaction term in the periodic boundary, depicted by a black oval, the colored spin is twisted by an angle $\phi_1$ (blue), $\phi_2$ (red) for the left-right and top-bottom  boundaries respectively.  Interaction terms which cross two boundaries get twisted by both phases, $\phi_3=\phi_1+\phi_2$ (pink) for the left-top boundary,  and $\phi_4=\phi_1-\phi_2$ (green) for the left-bottom one. The inner bounds are not depicted for clarity.}
	\label{fig:App}
\end{figure}
In this Appendix we show how TBC are implemented for the next-nearest-neighbors and chirality terms present in the model studied in sect~\ref{Sec:J1J2}. 
We show the scheme for both cases in Fig.~\ref{fig:App}. 
In the same way than in the next neighbors interactions (Fig.~\ref{fig:Lattice}) when a interaction term crosses the left-right (up-down) boundary the external spin gets twisted by a phase $\phi_1$, blue color ($\phi_2$, red color). The external spins in the top-left corner of the figures, are twisted by $\phi_3=\phi_1+\phi_2$ (pink color) because the interaction crosses both boundaries. In the next-nearest-neighbors case, there is, as well, an external spin in the bottom-left corner which crosses the left-right down-up border. Note that crossing the down-up border is the opposite as crossing the up-down one. Therefore, the spin in the bottom-left corner gets twisted by a phase $\phi_4=\phi_1-\phi_2$ (green color).
\end{document}